\documentclass[twocolumn,amssymb, amsmath, hyperref, aps, prd, showpacs]{revtex4}
\usepackage{bm}
\usepackage{graphicx}
\expandafter\ifx\csname package@font\endcsname\relax\else
 \expandafter\expandafter
 \expandafter\usepackage
 \expandafter\expandafter
 \expandafter{\csname package@font\endcsname}
 \fi
\def\sub#1{_{\mbox{\scriptsize{#1}}}}

\def\eg{\textit{e.g.} }

\def\re#1{(\ref{#1})}
\def\bkt#1{\left(#1\right)}

\newcommand{\mmm}{\medskip}
\newcommand{\bbb}{\bigskip}

\newcommand{\ff}{\phantom{t}}
\newcommand{\ii}{\textit}
\newcommand{\bb}{\textbf}

\newcommand{\no}{\noindent}

\newcommand{\be}{\begin{eqnarray}}
\newcommand{\ee}{\end{eqnarray}}
\newcommand{\nn}{\nonumber}
\newcommand{\bit}{\begin{itemize}}
\newcommand{\eit}{\end{itemize}}

\newcommand{\G}{\Gamma}

\def\Om{\Omega}

\def\pr{\prime}

\begin{document}
\title{Can we ever distinguish between quintessence and a cosmological constant?}
\author{Sirichai Chongchitnan}
\email{sc427@cam.ac.uk}
\author{George Efstathiou}
\email{gpe@ast.cam.ac.uk}
\affiliation{Institute of Astronomy. Madingley Road, Cambridge, CB3 OHA. United Kingdom.}

\begin{abstract}
Many ambitious experiments have been proposed to constrain dark energy and detect its evolution. At present, observational constraints are consistent with a cosmological constant and there is no firm evidence for any evolution in the dark energy equation of state $w$. In this paper, we pose the following question: suppose that future dark energy surveys constrain $w$ at low redshift to be consistent with $-1$ to a percent level accuracy, what are the implications for models of dynamical dark energy? We investigate this problem in a model-independent way by following quintessence field trajectories in `energy' phase-space. Attractor dynamics in this phase-space leads to two classes of acceptable models: 1) models with flat potentials, \ii{i.e.} an effective cosmological constant, and 2) models with potentials that suddenly flatten with a characteristic kink.
The prospect of further constraining the second class of models from distance measurements and fluctuation growth rates at low redshift ($z\lesssim3$) seems poor. However,  in some  models of this second class, the dark energy makes a significant contribution to the total energy density at high redshift. Such models can be further constrained from observation of the cosmic microwave background anisotropies and from primordial nucleosynthesis. It is possible, therefore, to construct models in which the dark energy at high redshift causes observable effects, even if future dark energy surveys constrain $w$ at low redshift to be consistent with $-1$ to high precision.

\end{abstract}

\date{May 2007}

\maketitle

\section{Introduction}
 
Ever since the discovery of cosmic acceleration \cite{riess98,perlmutter}, dynamical models of dark energy have been considered as serious alternatives to the cosmological constant (see \cite{copelandreview} for review). 
In the simplest dynamical models involving a scalar `quintessence' field \cite{ratrapeebles,wetterich,caldwell} rolling along a potential, the equation-of-state parameter $w=p/\rho$ is  time-dependent with $w>-1$. This contrasts with the cosmological constant, for which $w=-1$ at all times.

To test the nature of dark energy, a variety of astronomical techniques has been used to constrain $w$. We will not attempt to review the observational results in great detail here, but instead refer the reader to recent papers  \cite{essence2,wright,sahlen} that describe results from combinations of type Ia supernova (SNIa) surveys, cosmic microwave background  (CMB) anisotropies, galaxy redshift surveys and other cosmological data. These analyses are consistent with $w=-1$ to $\sim10\%$ accuracy, depending on exactly which datasets and theoretical assumptions (\eg curvature, evolution of $w$) are used. Although some data have suggested otherwise \cite{vishwakarma,alam}, there is a general consensus that the cosmological constant provides an excellent fit to the data and that there is no firm evidence for dynamical dark energy.

Nevertheless, understanding the nature of dark energy is of such importance for fundamental physics that a large number of ambitious new surveys have been proposed to tighten the limit on $w$ and its time evolution. These experiments include extensive ground and space-based surveys of distant type Ia supernovae \cite{snap,essence}, the Planck CMB  satellite \cite{planck}, large galaxy surveys to measure accurately the baryon acoustic oscillation (BAO) \cite{mcdonald,wfmos} over a wide range of redshift, X-ray and  Sunyaev-Zeldovich cluster surveys  \cite{haiman+,bartlett,rapetti} and ambitious ground and space-based weak lensing surveys \cite{heavens,dune}. We will not discuss the relative merits of these methods here. Summaries of many of these projects can be found in \cite{detf,eso+esa}.

Provided systematic errors can be kept under control, it may be possible to constrain $w$ to an accuracy of a percent or so within the next decade. Either we will find definitive evidence of a departure from $w=-1$, in which case dark energy must be dynamical, or the observational constraints will continue to tighten up around the cosmological constant value of $w=-1$. The problem that we wish to address in this paper is the following: if observations constrain $w$ to be very nearly $-1$ to high accuracy over some range of redshift, does this imply a cosmological constant or can we always construct radically different dynamical models that are consistent with the data?

In this paper, we will approach this problem in a model-independent way, without making specific assumptions about the form of the quintessence potential. While there are many quintessence models in the literature, some (loosely) inspired by string theory or supergravity, we clearly do not yet have a fundamental theory of dynamical dark energy. Given our current poor understanding of the physics of dark energy, quintessence models should be assessed  primarily on their ability to explain observations.

There have been many investigations of model-independent constraints on quintessence \cite{crittenden, huterer,sahlen,simpson,sahni+}. However, our approach, based on `energy' phase-space is quite different from these analyses. By investigating the attractor dynamics in this phase-space, we show in an intuitively clear way why it may well be impossible to ever distinguish between the cosmological constant and dynamically evolving models of quintessence.

\begin{table*}
\begin{tabular}{|c|c|c|c|c|}
\toprule
Attractor & $(x,y)$ coordinates & $\Om_q$ & $w_q$ & Condition on $\lambda$ \\
\colrule
\quad Quintessence-dominated \qquad & $\quad({\lambda/\sqrt{6}},\sqrt{1-\lambda^2/6})\quad$ & 1 & $\lambda^2/3-1$ & $\lambda<\sqrt{3(1+w_b)}$
\\
\quad Scaling solution & $(\sqrt{3/2\lambda^2}(1+w_b),\sqrt{3(1-w_b^2)/2\lambda^2})$ &\quad$3(1+w_b)/\lambda^2$\quad &$w_b$&  $\lambda>\sqrt{3(1+w_b)}$
\\
\botrule
\end{tabular}
\caption{Attractors in energy phase-space $x-y$ for the exponential potential $V(\phi)=V_0\exp^{-\lambda\kappa\phi}$. The values of the quintessence energy density $\Om_q$ and equation of state $w_q$ at each attractor are also shown. Given the conditions on $\lambda$ in the last column, the attractors are either stable nodes or spirals (see \cite{copeland} for detail). Here, the equation of state of the matter-radiation fluid, $w_b$, is $0$ during matter-dominated era, and $1/3$ during radiation era. }
\end{table*}

\section{Dynamics in the energy phase-space}

\subsection{Energy variables}

In this paper, quintessence potentials $V(\phi)$  are generated using the phase-space approach introduced in \cite{copeland} and subsequently used by a number of authors \cite{ng,bludman} to study the dynamics of specific quintessence potentials. To keep the discussion simple, we ignore any coupling of the quintessence field to matter, and we assume that the universe is spatially flat. In the phase-space approach, a quintessence model is represented by a trajectory in the plane of `energy' variables $(x,y)$ defined by: 
\be x\equiv {\kappa\dot{\phi}\over\sqrt{6}H}, \qquad y \equiv{\kappa\sqrt{V}\over\sqrt{3}H},\label{xy}\ee
where $\kappa=m\sub{pl}^{-1}=\sqrt{8\pi G}$ and $H={\dot a}/a$ is the Hubble parameter. These variables give the ratios of kinetic and potential energy of the quintessence field to the total energy density.  Energy conservation then requires:
\be x^2+y^2=1- \Omega\sub{m}-\Om\sub{r},\ee
where $\Omega\sub{m}$ and $\Om\sub{r}$ are the matter and radiation density parameters. This constrains the  quintessence trajectories to always lie within the unit circle in the energy phase-space. By restricting ourselves to $V(\phi)$ which are positive and monotonically decreasing, we only need to consider trajectories that lie within the first quadrant of the energy phase-space. These assumptions will not affect our main conclusions and are discussed further in Section \ref{discuss}.

Using the Friedmann equations, one can express the  trajectories as a function of redshift $z$:

\be{dx\over dz}&=&-{1\over 1+z}\bigg[-3x+\sqrt{3\over2}\lambda y^2+3x^3+\nn\\
&&\hskip 1 in {3\over2}x(1+w_b)(1-x^2-y^2)\bigg],\nn\\
{dy\over dz}&=&-{1\over 1+z}\bigg[-\sqrt{3\over2}\lambda xy + 3x^2y+\nn\\
&&\hskip 1 in{3\over2}y(1+w_b)(1-x^2-y^2)\bigg],\label{phase}\ee
where the equation of state of the matter-radiation fluid, $w_b$, is $0$ during matter-dominated era, and $1/3$ during radiation era. The `roll' parameter $\lambda$ given by \be\lambda\equiv -{1\over\kappa V}{dV\over d\phi},\label{lamb}\ee is analogous to the slow-roll parameter $\epsilon$ in inflation. In the next Section, we explain how $\lambda$ can be used to generate quintessence models stochastically.

\subsection{Attractor Dynamics}

If $\lambda$ takes a specific value $\lambda_0$, equation \re{lamb} integrates to give the exponential potential:
\be V= V_0 e^{-\kappa\lambda_0\phi}.\label{exp}\ee
Dark energy with the exponential potential has been considered by many authors, \eg\cite{wetterich,ratrapeebles,ferreira}. The attractor dynamics in energy phase-space has been thoroughly studied in \cite{copeland}. Table I summarises the locations and the conditions for existence of the attractors. Here we construct an arbitrary quintessence potential from a series of time-slices of exponential potential \re{exp} with varying $\lambda$ (see \cite{delamacorra,nastyman} for asymptotic analyses). The dynamical attractors are thus known exactly at every instant. Hence, as explained in detail in the next Section, a large number of trial quintessence  models may be generated using stochastic representation of the function $\lambda(z)$.

\section{Generating viable models of quintessence}\label{generate}

We shall now impose conditions on the phase-space trajectories motivated by present and possible future observational constraints. We emphasize that our goal here is to gain a physical understanding of quintessence models consistent with certain observational constraints. We therefore impose representative constraints from future experiments rather than modelling the experiments in great detail.

\subsection{Initial Conditions}

We first pick an initial point $(x_0,y_0)$ anywhere in the quadrant with uniform probability to start the trajectories at high redshift ($z\rightarrow\infty$). We introduce the change of variable 
\be z = {\zeta\over{1-\zeta}},\ee
which maps $z\in[0,\infty)$ to $\zeta\in[0,1)$. Thus,  initial conditions at arbitrarily early time can be set at $\zeta$ close to $1$.

We note that an advantage of the phase-space formalism is that a large (possibly infinite) range of initial conditions for quintessence corresponds to a small, finite region in the phase-space.

\subsection{Interpolating $\lambda(z)$}

From the initial condition $(x_0,y_0)$, trajectories can be evolved at high redshift according to equations \re{phase} provided the `roll' variable $\lambda(\zeta)$ is defined at every instant. To generate $\lambda(\zeta)$, we assign stochastic values to $\lambda$ in 100 uniformly distributed $\zeta-$bins in the range $0.75\leq\zeta<1$. Values at intermediate $\zeta$ are obtained by linear interpolation between bins. 

In each $\zeta$-bin, we draw $\lambda$ from a uniform distribution within the range
\be 0\leq\lambda\leq10.\label{lambrange}\ee
Larger values of $\lambda$ correspond to models with steeper potentials. Because critical behaviour of the attractors only occurs around $\lambda=2$ and $\sqrt{3}$, the range \re{lambrange} is general enough for us to identify a representative behaviour of viable quintessence models. A more general scenario is where $\lambda$ changes sign, corresponding to potentials that are not monotonically decreasing. In Section \ref{discuss}, we explain why our conclusions also extend to potentials of this type.

\subsection{Evolution of trajectories}\label{evolution}

Trajectories are evolved via Equations \re{phase} using the $\zeta$ variable until a low redshift ($z=3$) where we revert to $z$ as the time variable to analyse observables over a finer range of bins at low redshift. In $z\in[0,3]$, we make a further set of 25 redshift bins between which $\lambda(z)$ is interpolated as described above.

We are interested in the class of models that are compatible with tight observational constraints expected from future experiments as described in the introduction. Let $z\sub{obs}$ denote the redshift  at which observational constraints will be tightest. The actual value of $z\sub{obs}$ will depend on the range of techniques used to establish the constraints. For example, $z\sub{obs}\sim0.3$ for supernova surveys and $\sim0.6$ for future weak lensing surveys \cite{simpson}. For this paper, we will adopt $z\sub{obs}=0.3$ and note that qualitatively, we can recalibrate to a different value of $z\sub{obs}$ by applying an additive constant to the redshift. Here, we adopt the following constraints: 

\begin{enumerate}
\item[(a)] the dark energy equation of state $w_q$ is  tightly constrained at low redshift to be close to $-1$ to a percent accuracy at $z=z\sub{obs}$:
\be -1\leq w_q\leq -0.99 \quad\mbox{at }z=z\sub{obs}=0.3,\label{c}\ee
where \be w_q = {x^2-y^2 \over x^2+y^2}.\label{wq}\ee

\item[(b)] the dark energy density $\Om_q$ will be tightly constrained at low redshift to a percent accuracy (compatible with the current bounds on $\Om_q$ from SNIa and CMB experiments \cite{wright,essence2}):
\be  0.73\leq\Om_q\leq0.74\quad\mbox{at }z=0,\label{d}\ee where \be \Om_q = {x^2+y^2}\label{oq}.\ee

\end{enumerate}

\begin{figure}
\includegraphics[width=8cm, height=8.5cm]{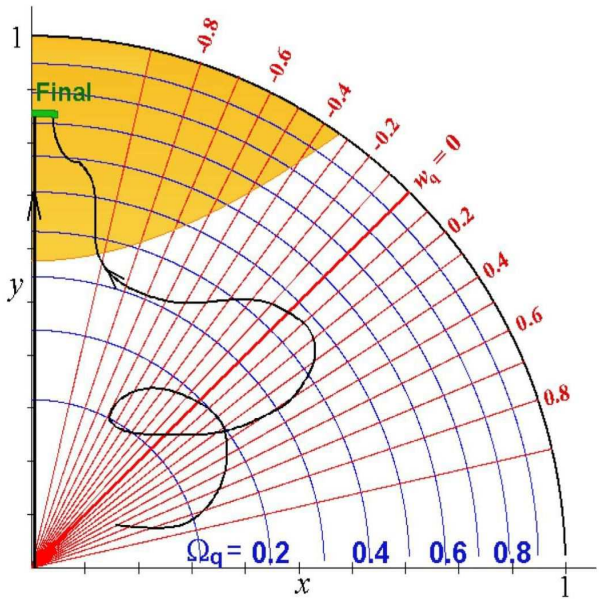}
\caption{Energy phase-space (see Equation \re{xy}). A viable quintessence model can be represented by a trajectory that evolve into a small region defined by the imposed conditions (Eq. \re{c} and \re{d}). Each point $(x,y)$ corresponds to the observables $(w_q,\Om_q)$ via relations \re{wq} and \re{oq}, shown here as circular and radial contours. Shaded region denotes where cosmic acceleration occurs ($w\sub{eff}=w_q\Om_q<-1/3$). A trajectory may mimic the cosmological constant by running closely along the $y$-axis, or more generally, it may be a complicated curved path around the quadrant.}
\label{fig1}
\end{figure}

Constraints (a) and (b) define a very small region in  phase-space through which trajectories must pass $z=z\sub{obs}$ and terminate at $z=0$. Thus, the problem of finding viable quintessence potentials translates to finding those trajectories that evolve into a small region defined by the constraints (a)-(b). This is illustrated in Figure \ref{fig1}. Trajectories may be straight lines  (as in the case of the cosmological constant), or complicated paths around the quadrant.


We ran a code that generates a large number of quintessence models and selected those that satisfy the criteria described in this Section. In this way, we determine the generic type of quintessence potentials which may be allowed by future dark energy experiments if the observations continue to favour $w=-1$.

\subsection{Acceptable models}\label{accept}

\begin{figure*}
\includegraphics[height=6cm]{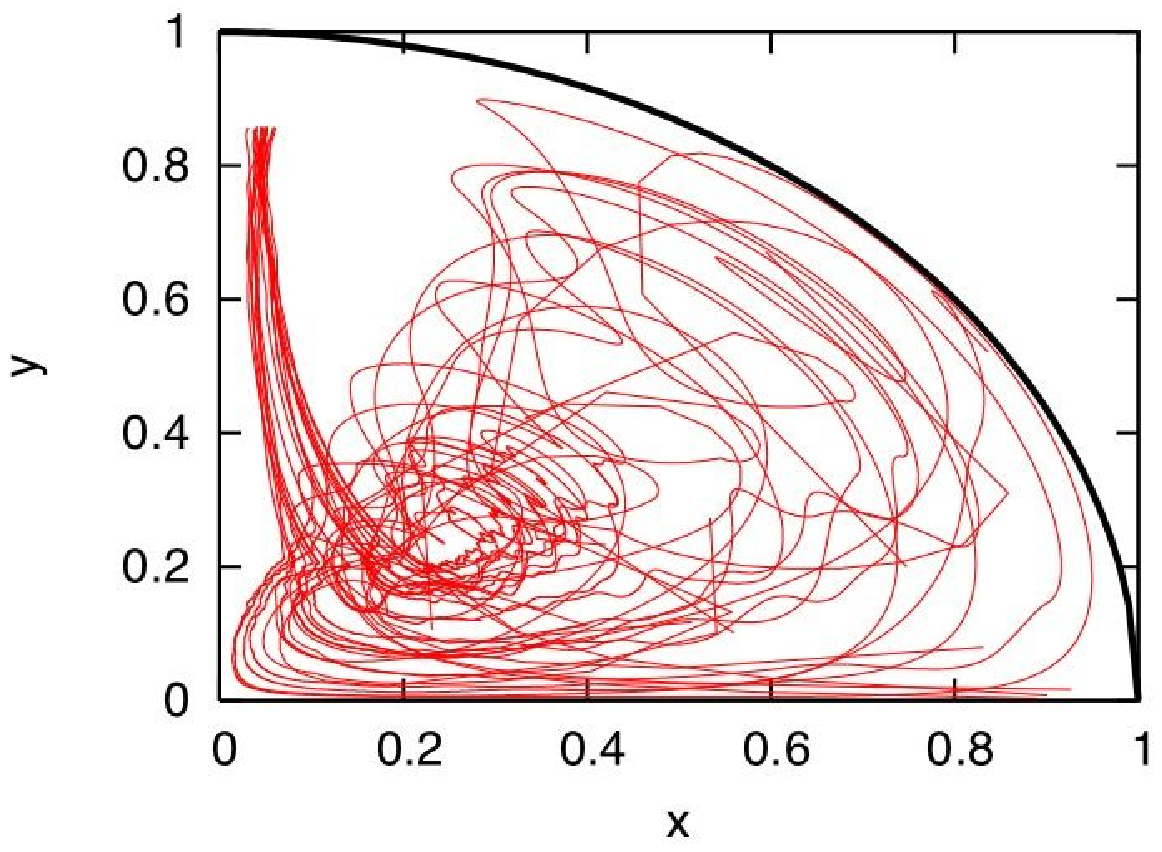}
\includegraphics[height=5cm,width=12cm]{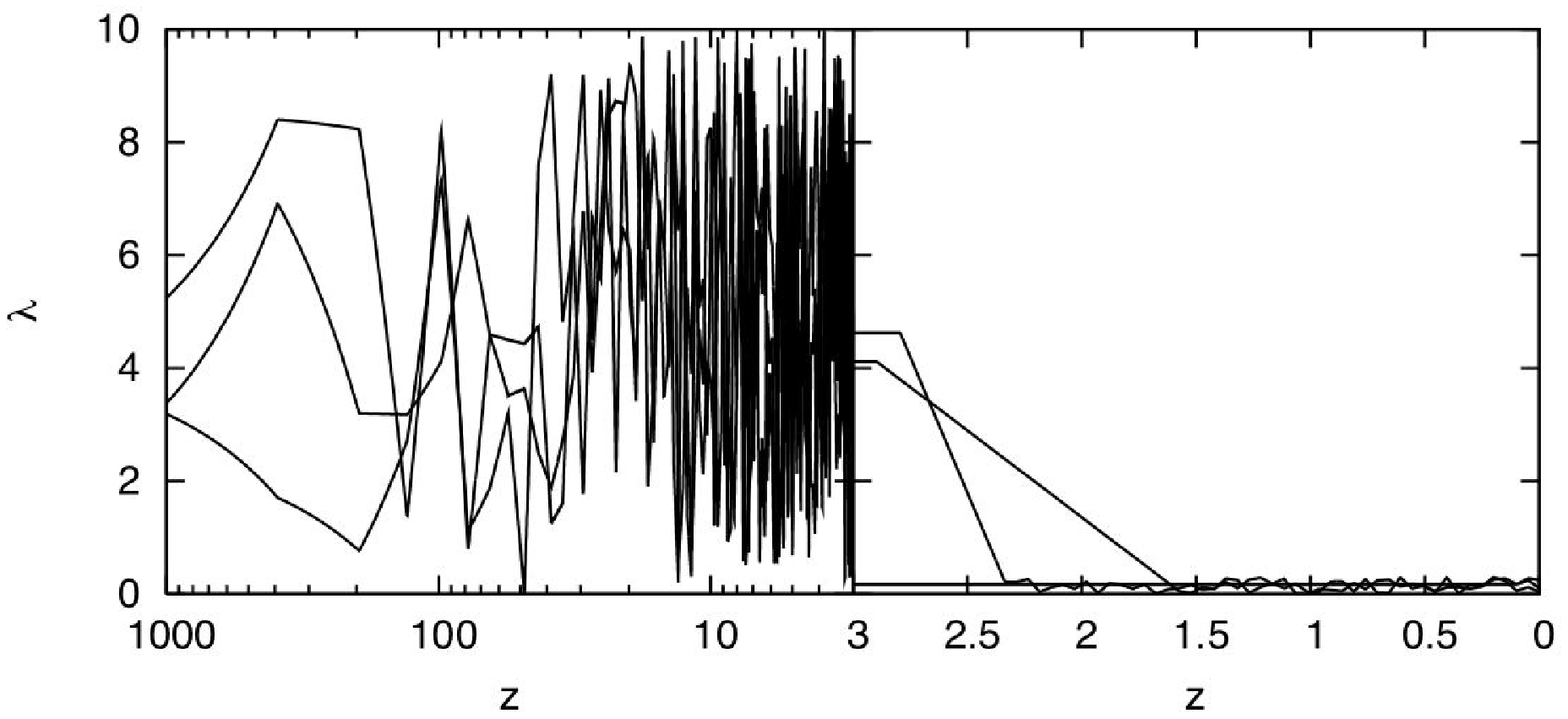}
\includegraphics[height=5cm,width=12cm]{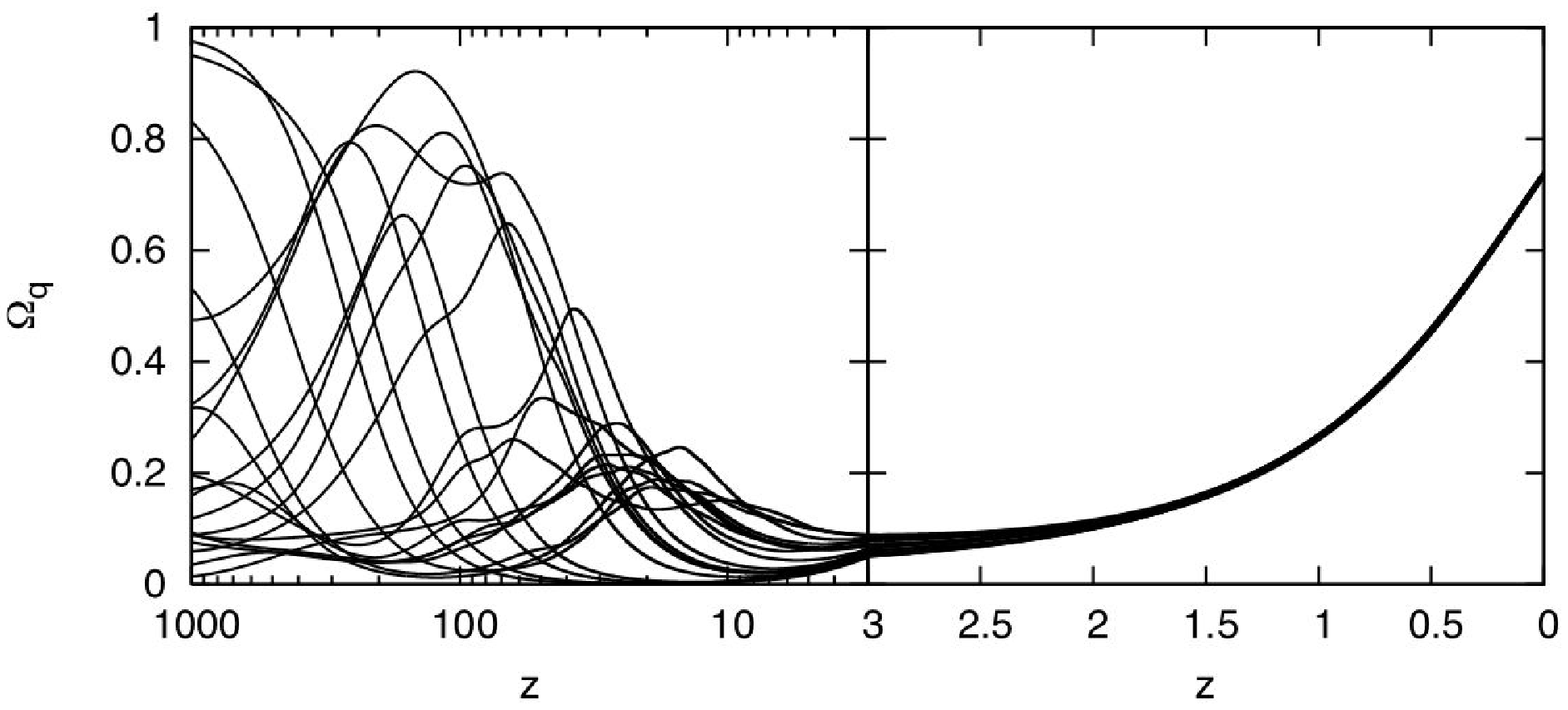}
\includegraphics[height=5cm,width=12cm]{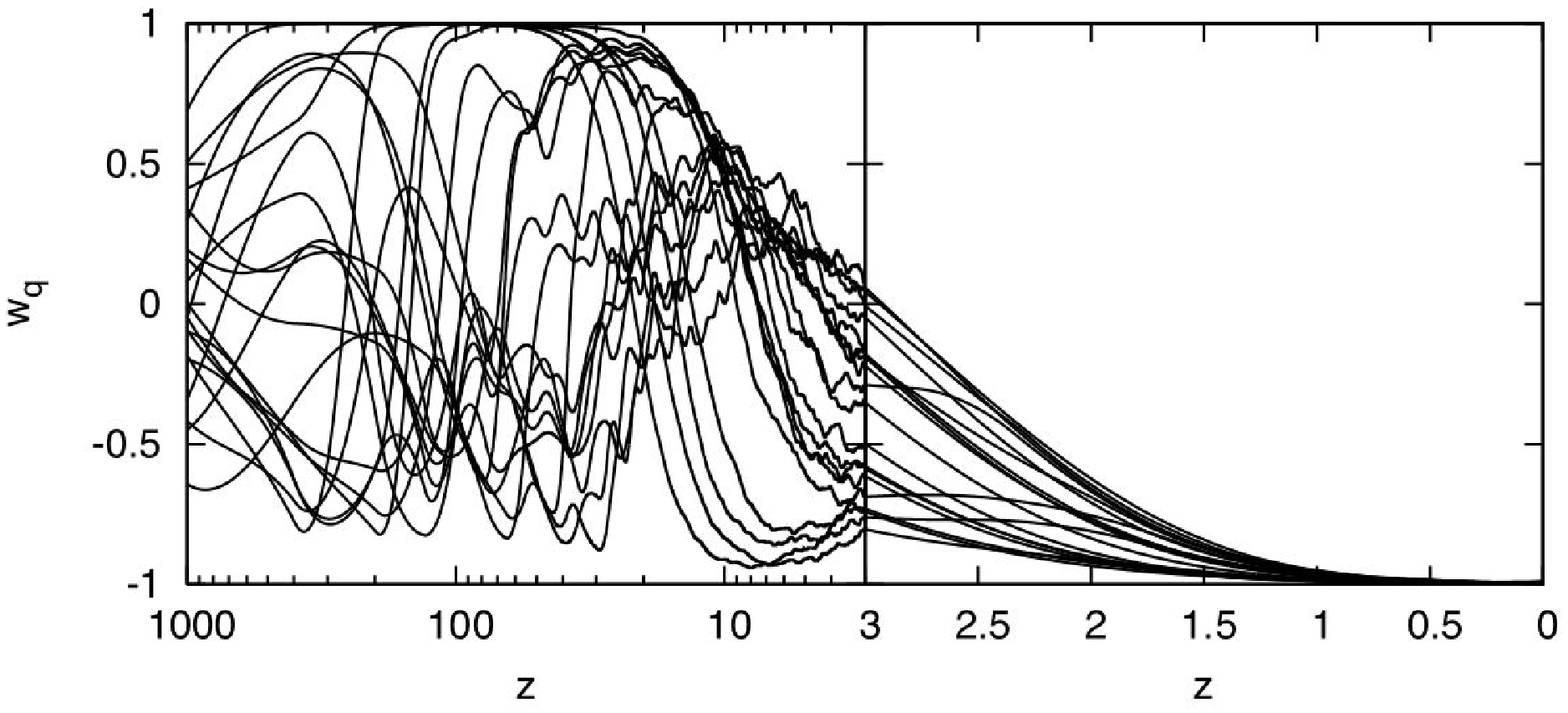}
\caption{Examples of trajectories of quintessence models that satisfy the imposed constraints \re{c} and \re{d} (top panel). From a wide range of initial conditions, they approach the scaling solution at early time. But once the scaling solution decays, they subsequently evolve towards the region defined by Eq. \re{c}-\re{d}. As explained in the text, these trajectories are generated by stochastic representation of $\lambda(z)$, shown in the second panel (for clarity we have shown only 3 such representations in this panel). The lower panels show the variations of the dark energy density $\Om_q(z)$ and its equation of state $w_q(z)$, both of which show strong evolution at high redshift.}
\label{fig2}
\end{figure*}

The trajectories fall into two main categories: 1) static and `skater' models associated with flat potentials, and 2) dynamical models with trajectories that veer around the phase-space in a particular way before terminating. We now describe the dynamics of each type of models in detail.

\subsubsection{Static and skater models}

These models are associated with flat potentials, with  $\lambda\simeq0 $ throughout their evolutionary histories. Static models have $x\simeq0$ and behave like a cosmological constant. For `skater' models \cite{sahlen2}, the quintessence field evolves along a flat potential with a constant non-zero kinetic energy. Hence, the phase-space trajectory of   skater models is the line $y=kx$, where $k=\sqrt{2V}/\dot{\phi}$.

 Starting from a generic initial condition within the quadrant and given that $\lambda\ll 1$, the trajectories   quickly evolve towards the quintessence-dominated attractor $(x,y)\approx(0,1)$. To produce static or skater models that satisfy our criteria (a) and (b), we found it necessary to set the initial coordinates to very precise values which were found by backward integration. This simply reflects the fine-tuning of the amplitude of the potential if dark energy is the cosmological constant. For skater models, the constraint (a) on $w_q$ translates to an additional tuning of the kinetic energy such that \be |\dot\phi|<\sqrt{2V\over199},\ee
and so they behave almost identically to the static models. 


\subsubsection{Dynamical models}\label{main}

The main purpose of this paper is to determine what types of dynamical quintessence models will satisfy our imposed constraints and to assess whether these are observationally distinguishable from the static models. Successful dynamical models evolve in one of the following ways: 

\begin{enumerate}
\item[(i)] The trajectory evolves haphazardly throughout its history, avoiding the attractors in Table I.

\item[(ii)] The trajectory evolves towards the scaling-solution attractor, which subsequently decays.

\item[(iii)] If $\lambda\lesssim1$, the trajectory  evolves from a quintessence-dominated regime at high redshift to satisfy the observational constraints at $z\sub{obs}$. 

\end{enumerate}

It is very difficult to produce type (i) trajectories that satisfy the imposed constraints (a) and (b). This is because the scaling solution is a global attractor \cite{wands}. Trajectories of type (iii) are unacceptable because an extended period of quintessence domination at high redshift would have disrupted structure formation.  Trajectories of type (ii) can produce acceptable models of dynamical dark energy, which we now discuss in more detail.

To satisfy our stringent constraints that $w_q$ is close to $-1$ at $z\sub{obs}$, clearly all solutions must behave like the cosmological constant at low redshift. In the type (ii) solutions, this behaviour is matched to a scaling solution at high redshift, leading to interesting dynamical behaviour. Since such a scaling solution has $\Om_q=3(1+w^2)/\lambda^2$ and necessarily $\lambda$ must be small at low redshift, an acceptable model requires a rapid transition of $\lambda$ if the universe is to be matter-dominated at high redshift. The potentials corresponding to these models therefore have a sharp transition from a steep to a shallow slope at some characteristic field value.  

Figure \ref{fig2} shows phase-space trajectories and the corresponding forms of $\lambda$, $\Om_q$ and $w_q$ for   dynamical models that satisfy the imposed constraints.   The scaling solution attracts trajectories from a wide range of initial conditions, unlike the static solutions. By construction, all of these models have $w_q\approx-1$ at $z\lesssim1$. However, unlike tracker models considered in the literature, in which $w_q$ varies slowly with redshift \cite{steinhardt}, many models shown in Figure \ref{fig2} display very strong evolution in $w_q$. The evolution of $w_q$ for these models is poorly described by parametrizations such as $w=w_0+w_az(1+z)^{-1}$, often used in the literature, \eg \cite{wright,huterer} and used by the dark energy task   force to define a figure of merit for comparing future dark energy surveys \cite{detf}. Note that $w_q$ for these models varies wildly above $z\sim5$. At $z\lesssim3$, all of the models show very similar evolution in $\Om_q$ though, as with $w_q$, $\Om_q$ becomes unconstrained at high redshift. 

\begin{figure*}
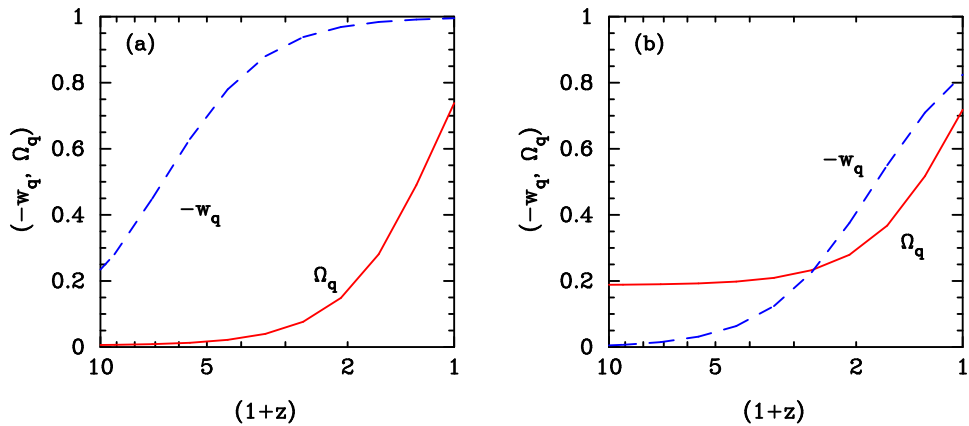

\includegraphics[height=6cm,  angle = -90]{qfigs3a.ps}
\qquad \includegraphics[height=6cm,  angle = -90]{qfigs3b.ps}

\caption{Evolution of $w_q$ and $\Omega_q$ with redshift for the double
  exponential potential of Equation \re{DE1}. The parameters have been
set to  $V_1 = 7\times 10^{-111}m\sub{pl}^4$, $V_2 = 4 \times
10^{-121}m\sub{pl}^4$, $\lambda_2 = 0.05$ with $\lambda_1 = 25$ in panel (a) 
and $\lambda_1=4$ in panel (b). }
\label{fig3}
\end{figure*}

To summarise, we have investigated the implications if  future observations continue to tighten the constraint on $w_q$ around $-1$ with increasingly high precision. Even if future observations were to achieve $1\%$ accuracy in $w_q$ at low redshift, our results show that   two classes of models would be compatible with the constraints: static (skater) models that are indistinguishable from a cosmological constant, and a class of dynamical models for which $w_q$ makes a transition from a scaling solution to $w_q=-1$ at recent epochs. No matter how tight the constraint on $w_q$ at low redshift, it is always possible to find dynamical dark energy models of the second class, which shows strong evolution at high redshift. This is the main result of this paper.

\section{Relation to the double exponential potential}

Some of the results of the previous Section can be understood by
considering the double exponential potential \cite{barreiro}
\begin{equation}
V(\phi) = V_1{\rm e}^{-\lambda_1 \kappa \phi}+ V_2{\rm e}^{-\lambda_2 \kappa \phi}. \label{DE1}
\end{equation}
As discussed above, a pure
exponential potential has a late time attractor with $w_q = w_b$,
$\Omega_q = 3(1+w_b)/\lambda^2$ if $\lambda^2 > 3(1 + w_b)$ and an
attractor with $w_q = - 1 + \lambda^2/3$, $\Omega_q=1$ if $\lambda^2 <
3(1 + w_b)$. Evidently, it is not possible, using a single exponential
potential, to construct a model with a matter/radiation dominated
phase at high redshift with a transition to a quintessence dominated
phase at low redshift. It is, however, easy to construct such models
with a double exponential potential (\ref{DE1}) if $\lambda_1\gg
\lambda_2$.  Figure \ref{fig3} shows two examples. In Figure \ref{fig3}a, we set
$\lambda_1 = 25$, $\lambda_2 = 0.05$, $V_1 = 7\times 10^{-111}m\sub{pl}^4$,
$V_2 = 4 \times 10^{-121}m\sub{pl}^4$. These parameters easily satisfy the
constraint $w_q < -0.99$ at low redshift, and the high value of
$\lambda_1$ ensures that the universe is matter/radiation dominated at
high redshift. The potential used to generate Figure \ref{fig3}b has identical
values for $V_1$, $V_2$ and $\lambda_2$, but has $\lambda_1 =4$. The
lower value of $\lambda_1$ leads to a much stronger evolution of $w_q$
with redshift but fails to match the target value of $w_q < -0.99$ by
the present day by a wide margin. Although the double exponential
potential shows some of the features of the models described in the
previous Section, it is not flexible enough to produce $w_q < -0.99$
at low redshift {\it and} rapid evolution of $w_q$. The numerical
technique described, however, allows us to find acceptable models
with both fast and slow evolution of $w_q$.

\section{Distinguishing between static and dynamical models}\label{dist}

\begin{figure}
\centering
\includegraphics[width=8.5cm]{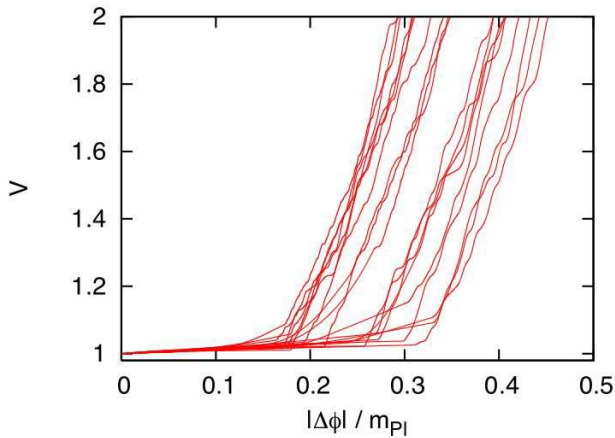}
\caption{The potential $V$ as a function of $|\Delta\phi|$, the absolute change in the quintessence field value with $\phi=0$ today. These potentials correspond to those quintessence trajectories shown in figure \ref{fig2}. The steep parts can be pushed to arbitrarily high field values. The prospect of distinguishing these models from the cosmological constant is discussed in the text.}
\label{fig4}
\end{figure}

The quintessence potential $V(\phi)$ can be recovered from the energy variables by first integrating Equation \re{xy} to find the quintessence field value as a function of $z$:
\be \phi(z) = -\sqrt{6}\int_0^z {x(z^\prime)\over1+z^\prime}dz^\prime,\label{phiz}\ee
where $\phi$ is in unit of reduced Planck mass $m\sub{pl}$ and is arbitrarily set to be zero at $z=0$. Next, we integrate equation \re{lamb} to find $V(z)$:
\be V(z) = V_0\exp\bkt{\sqrt{6}\int_0^z{\lambda(z^\prime) x(z^\prime)\over 1+ z^\prime}dz^\prime},\label{vz}\ee
where $V_0$ is the value of the effective cosmological constant today. Combining \re{phiz} and \re{vz} gives the potentials $V(\phi)$ as shown in Figure \ref{fig4} for the models plotted in Figure \ref{fig2}.

The key feature of these potentials is the sudden flattening at field values characteristic of the transition to quintessence domination at low redshift. These models can only be distinguished from the cosmological constant by observational tests that are sensitive to the steep part of the potential at high redshift. Unfortunately, the number of feasible  observational tests is limited, amounting to estimates of distance (either angular diameter or luminosity distance) or to measurements of the growth rates of fluctuations.

\subsection{Distance measurements}

Luminosity distance $d_L$ is given as a function of redshift by

\be  d_L(z)= {c\over H_0}(1+z)\times\hskip 2in \\
\int_0^{z}{dz^\prime\over\sqrt{\Om_m(1+z^\prime)^3+\Om_r(1+z^\prime)^4+\Om_q(1+z^\prime)^3e^{W(z^\prime)}}},\nn\ee

\be W(z)= 3\int_0^{z}{w_q(z^\prime)\over 1+z^\prime}dz^\prime.\ee

The angular diameter distance $d_A$ is related to $d_L$ by \be d_A = {d_L\over{(1+z)^2}}.\ee

\begin{figure*}
\includegraphics[width=9cm,height = 5.82cm]{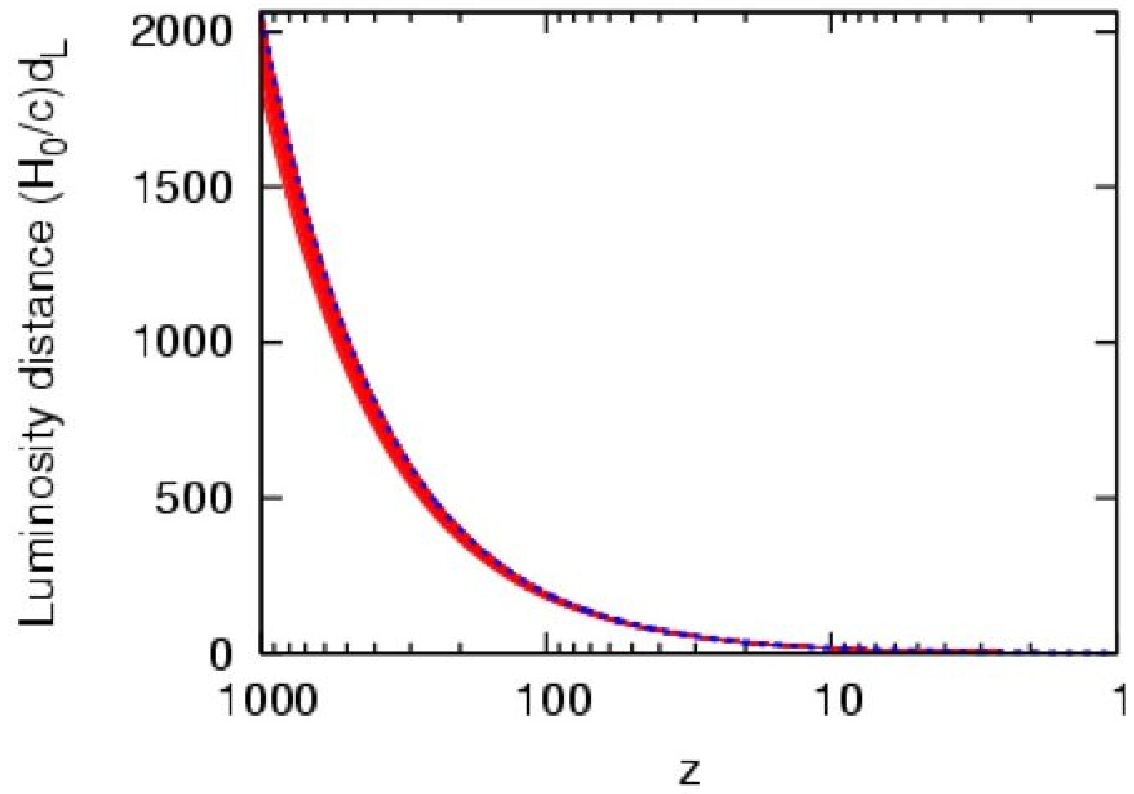}
\includegraphics[width=8cm, height = 5.82cm]{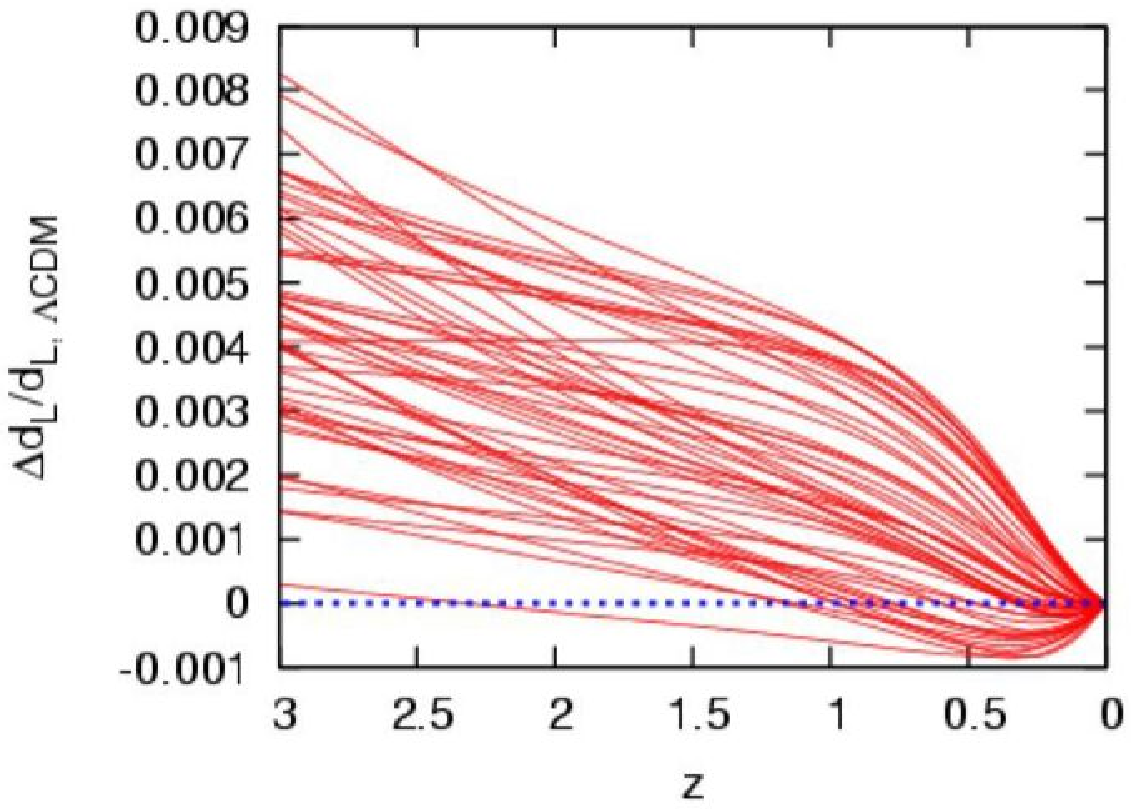}
\caption{Evolution of the dimensionless luminosity distance $(H_0/c)d_L(z)$ in the redshift range 0 to 1000 for quintessence models generated by the $\lambda$ parametrization as shown in Figure \ref{fig4}. Dashed line shows the corresponding values for the $\Lambda CDM$ model. The fractional change with respect to the $\Lambda CDM$ model (Equation \re{fracdl}) is shown in the right-hand panel.  }
\label{fig5}
\end{figure*}

Figure \ref{fig5} shows the luminosity distance as a function of redshift for the models shown in Figure \ref{fig4}. The panel to the right shows the fractional change in the luminosity distance with respect to a constant $\Lambda$ model:
\be {\Delta d_L \over d_L}={d_L(z)-d_{L, \Lambda CDM}(z) \over d_{L, \Lambda CDM}(z)},\label{fracdl}\ee 
at low redshift on an expanded scale. One can see that over this redshift range, the typical fractional changes are of order half a percent. Such small variations in distance would be difficult to constrain by either baryon oscillation surveys or large surveys of high-redshift type-Ia supernovae. For example, if we ignore any systematic effects, the fractional error in distance from a sample of $N$ supernovae at redshift $z$ with intrinsic magnitude dispersion $\sigma_i\sim0.1$ is

\be {\Delta d_L \over d_L}={\ln10\over5}{\sigma_i\over \sqrt{N}},\ee
and so typically $10^3$ or more supernovae in each of several redshift bins would be required to constrain the models of Figure \ref{fig5}. Achieving better than a percent accuracy in distance measurements via baryon oscillation surveys would be formidably difficult even if systematic errors associated with nonlinearities and biasing were well-understood \cite{glazebrook}. It seems unlikely, therefore, that distance measurements could strongly constrain the models of figure \ref{fig5}.

\begin{figure*}
\includegraphics[width=9cm]{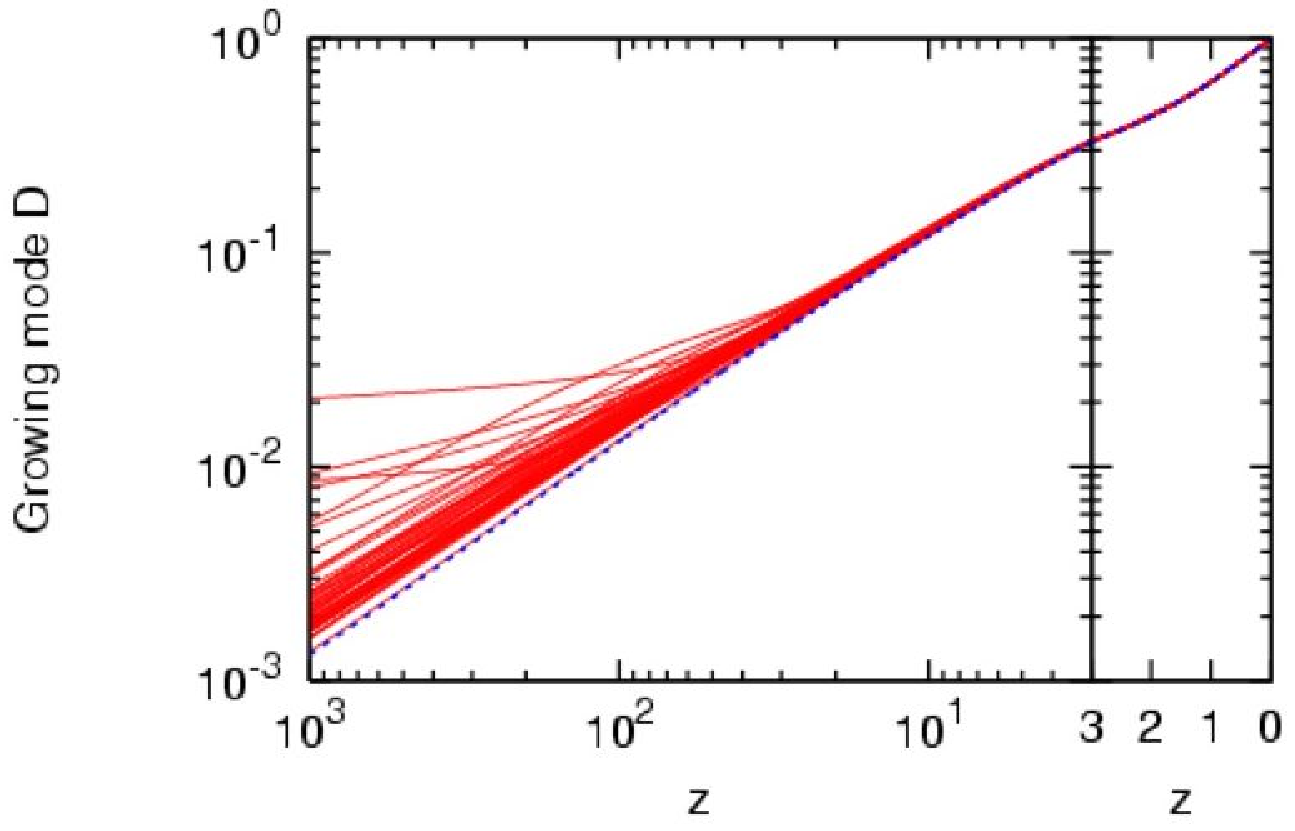}
\includegraphics[width=8cm, height = 5.72cm]{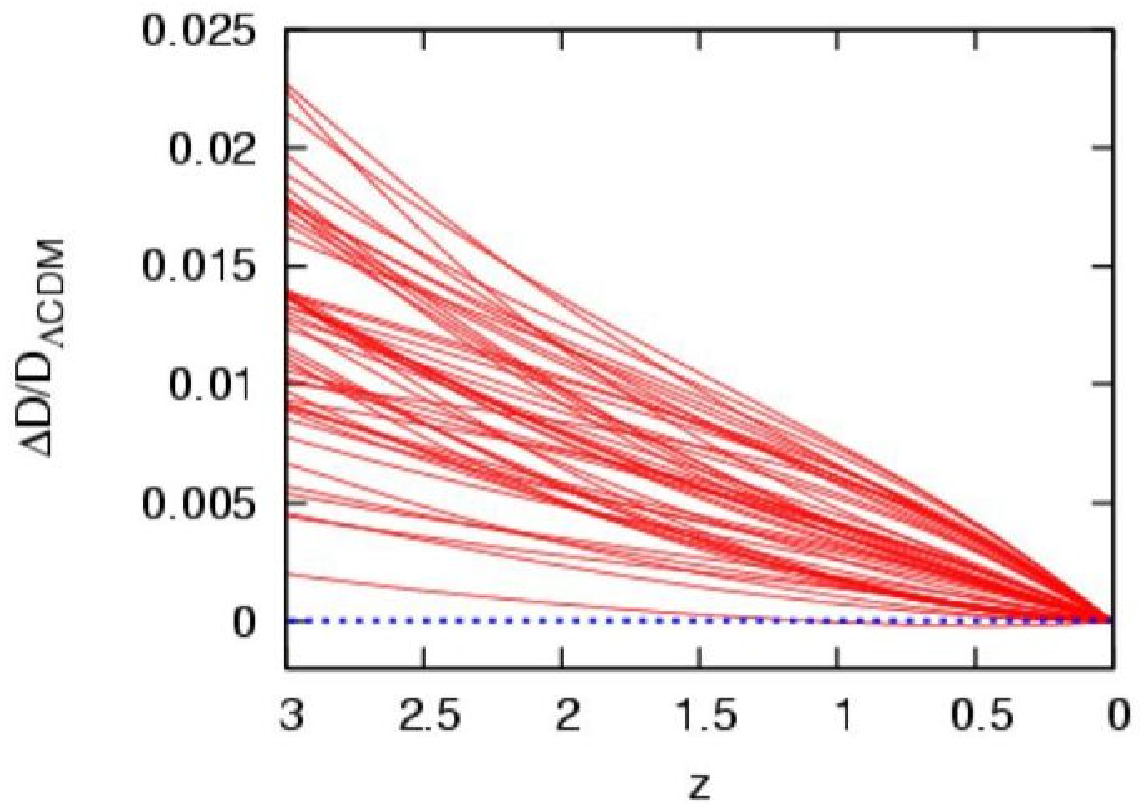}
\caption{Evolution of the magnitude of growing mode of density perturbation $D(z)$ in the redshift range 0 to 1000 for quintessence models shown in Figure \ref{fig4}, with normalization $D(0)=1$. Dashed line shows the corresponding values for the $\Lambda CDM$ model. The fractional change with respect to the $\Lambda$CDM model (Equation \re{fracD}) is shown in the right-hand panel. }
\label{fig6}
\end{figure*}

\subsection{Growth rates}

Growth of large-scale structures can be used to constrain the dark energy equation of state \cite{linder,wang,benabed,doran}. Let $D$ denote the amplitude of the growing mode of  density perturbations. The growth suppression parameter $f(z)$ is defined as
\be f = {d\ln D\over d\ln a}.\ee

The redshift-evolution of $f$ is given by  \cite{wang}

\be {df\over dz}={1\over 1+z}\bkt{f^2+{f\over2}(1-3w_q\Om_q)-{3\over2}\Omega_m},\ee
where $f$ is normalized to be 0 well into the radiation-  dominated era. $D$ is then given by
\be D(z)=\exp\bkt{-\int_0^z{f(z^\prime)\over 1+z^\prime}dz^\prime},\ee with the normalization $D(0)=1$. Figure \ref{fig6} shows the evolution of $D(z)$ for the models of figures \ref{fig4} and \ref{fig5}. The right-hand panel shows the fractional error with respect to the growth rate of the constant $\Lambda$ model:

\be {\Delta D \over D}={D(z)-D_{\Lambda CDM}(z) \over D_{\Lambda CDM}(z)},\label{fracD}\ee on an expanded scale for low redshift. The variation in growth rates is of order a percent or so at low redshift, and would be difficult to constrain via observations of the mass function of rich clusters \cite{linder} or the integrated Sachs-Wolfe effect in the CMB  \cite{corasaniti,giannantonio}.

In some of the models, the growth rates vary quite substantially from the constant $\Lambda$ model at high redshift. As discussed in Section \ref{accept}, this can occur if the quintessence potential has a gentle gradient, allowing solutions in which the dark energy makes a significant contribution to the total energy density at high redshift. Constraints on this type of behaviour arise from the consistency of primordial nucleosynthesis (limiting the contribution of dark energy so that the appropriate expansion rate at the nucleosynthesis epoch is maintained \cite{corasaniti2}). Additional constraints on the dark energy contribution at around the recombination era can be derived from the primary CMB anisotropies \cite{wright,bean}, and in particular, from the consistency of the amplitude of the fluctuations at low redshift \eg \cite{kunz}. Nevertheless, it is clear from Figure \ref{fig6} that it is easy to construct  acceptable dynamical models in which the dark energy is always subdominant at high redshift.

\section{Discussion and conclusion}\label{discuss}

\subsection{Non-monotonic potentials}
The space of models opens up if we consider quintessence in which the field may also roll uphill. In this case, trajectories can evolve into the $x<0$ quadrant of the energy phase-space. Consequently, trajectories that satisfy the imposed constraints \re{c}-\re{d} simply alternate between the scaling solution in the $x>0$ quadrant, and that in the $x<0$ quadrant whenever the roll direction changes. This gives rise to models which behave like the cosmological constant even at high redshift, with a fluctuating $\lambda(z)$ of high amplitude. Distinguishing between such models and the cosmological constant is even more difficult than the monotonic potentials discussed above. Thus, we have not considered non-monotonic potentials in detail here. For specific examples, see \cite{albrecht,skordis}.

\subsection{Polynomial potentials and the `flow' approach}

A number of authors have investigated observational constraints on simple low-order polynomial (or Pad\'{e} approximate) parametrizations of the quintessence potential \cite{sahlen2,sahlen}. In a recent paper \cite{huterer}, the inflationary flow approach of \cite{hoffman,kinney} was adapted to study quintessence, but  this formulation is equivalent to adopting a low-order polynomial for the quintessence potential.  

Such specific parametrizations, while useful, cannot easily produce the potentials with sharp kinks shown in figure \ref{fig4}. To illustrate this point, we supplemented the differential equations \re{phase} by a set of `quintessence flow equations':
\be{d\lambda\over dz}&=&{1\over 1+z}\sqrt{6}x\lambda^2(\G_1-1),\\
{d\G_n\over dz}&=&{1\over 1+z}\sqrt{6}x\lambda\G_n\bkt{1-\G_1-\G_n+\G_{n+1}},\ff n\geq1\nn\ee  which can be truncated at an arbitrarily high order. Here:
\be\G_n\equiv{V^{(n+1)}V\over V^{(n)}V^\pr},\ee
is the generalised curvature parameter, with $\G_1$ reducing to the usual definition of $\G \equiv{VV^{\pr\pr}/V^{\pr2}}$
usually found in analysis of tracker potentials. With a low value of $n$, these equations cannot produce models with a sharp kink in the potential. On the other hand, for higher values of $n$, the initial conditions and $\G_n$ parameters need to be finely tuned to generate models similar to those in Figure \ref{fig4}.
In contrast, it is straightforward to construct dynamical models that satisfy our imposed constraints using the $\lambda$ parametrization discussed in Section \ref{generate}.

\subsection{Implications}

We have presented a study of quintessence by considering  stochastic trajectories in energy phase-space (Figure \ref{fig1}). This formalism allows us to generate positive, monotonic potentials that are consistent with  imposed constraints on the dark energy density and its equation of state.

As an illustration of the approach, we have investigated the implications for dynamical models of dark energy should future ambitious observational projects succeed in constraining $w$ to be $-1$ at low redshift to an accuracy of a percent.

One set of models that can satisfy these constraints is  obviously the cosmological constant or `skater' models with extremely flat potentials and a slowly-moving field. However, we also find a second class of acceptable model in which the potentials have a sharp transition from a steep to a shallow slope as shown in figure \ref{fig4}. For this class, the field follows a scaling solution at high redshift, and so the dynamics is insensitive to the initial conditions. Although $w_q$ for this type of model can evolve strongly with redshift , the deviations in distance measures and linear growth rates compared to a constant $\Lambda$ model over the redshift range $0\lesssim z\lesssim3$ are so small that it will be extremely difficult to rule them out observationally \footnote{This is assuming that  systematic errors in future astronomical datasets are well-understood (see \cite{krauss,davis,hsiao} for recent discussions).}

Evidently, it is impossible ever to exclude non-trivial dynamical models of dark energy no matter how precise the observational constraints at low redshift. If, as assumed here, the observational constraints on $w$ continue to tighten up around $-1$ to high accuracy, then in a Bayesian-Information-Criterion sense, a cosmological constant will be favoured over dark energy models involving more parameters \cite{sahlen}. Nevertheless, the numerical scheme described here shows that it is straightforward to construct dynamical models which can satisfy observational constraints to an arbitrary precision at low redshift, and yet show interesting dynamical behaviour at high redshift. Our only handle on this type of model comes from constraints on the dark energy density at high redshift (for example, from limits on the growth rates of fluctuations \cite{wright,bean}). Even if the heroic efforts of the next generation of dark energy surveys succeed in constraining $w=-1$ to exquisite accuracy, we must remain alert to the possibility that dark energy may have been dynamically important at high redshift. 

\bbb

\no\centerline{\bb{Acknowledgement}}

\mmm

S.C. is grateful for communication with Lindsay King, David Wands and Jeremy Sakstein. He also thanks the support of a Dorothy Hodgkin scholarship from PPARC. Our work has been supported by PPARC.

\bibliographystyle{unsrt}
\bibliography{quint}

\end{document}